\documentclass[journal=ancac3,manuscript=article]{achemso}

\usepackage[version=3]{mhchem} 
\usepackage{braket,comment}

\title{Towards Engineering Intrinsic Linewidths and Line-Broadening in Perovskite Nanoplatelets}

\author{Albert Liu}

\affiliation{Department of Physics, University of Michigan, Ann Arbor, Michigan, USA}

\author{Gabriel Nagamine}
\affiliation{Instituto de Fisica, Universidade Estadual de Campinas, Campinas, Sao Paulo, Brazil}

\author{Luiz G. Bonato}
\affiliation{Instituto de Quimica, Universidade Estadual de Campinas, Campinas, Sao Paulo, Brazil}

\author{Diogo B. Almeida}
\affiliation{Instituto de Fisica, Universidade Estadual de Campinas, Campinas, Sao Paulo, Brazil}

\author{Luiz F. Zagonel} 
\affiliation{Instituto de Fisica, Universidade Estadual de Campinas, Campinas, Sao Paulo, Brazil}

\author{Ana F. Nogueira} 
\affiliation{Instituto de Quimica, Universidade Estadual de Campinas, Campinas, Sao Paulo, Brazil}

\author{Lazaro A. Padilha}
\email{padilha@ifi.unicamp.br}
\affiliation{Instituto de Fisica, Universidade Estadual de Campinas, Campinas, Sao Paulo, Brazil}

\author{Steven T. Cundiff}
\email{cundiff@umich.edu}
\affiliation{Department of Physics, University of Michigan, Ann Arbor, Michigan, USA}

\begin{document}

\begin{abstract}
    Perovskite nanoplatelets possess extremely narrow absorption and emission linewidths, which are crucial characteristics for many optical applications. However, their underlying intrinsic and extrinsic line-broadening mechanisms are poorly understood. Here, we apply multi-dimensional coherent spectroscopy to determine the homogeneous line-broadening of colloidal perovskite nanoplatelet ensembles. We demonstrate control of not only their intrinsic linewidths, but also control of various broadening mechanisms by tuning the platelet geometry. Remarkably, we find that decreasing nanoplatelet thickness by a single polyhedral layer results in a 2-fold reduction of the inhomogeneous linewidth and a 3-fold reduction of the intrinsic homogeneous linewidth to the sub-meV regime. In addition, our measurements suggest homogeneously broadened exciton resonances in 3-layer (but not necessarily 4-layer) nanoplatelets at room-temperature.
    
    \begin{tocentry}
        \includegraphics[width=9cm,height=5cm]{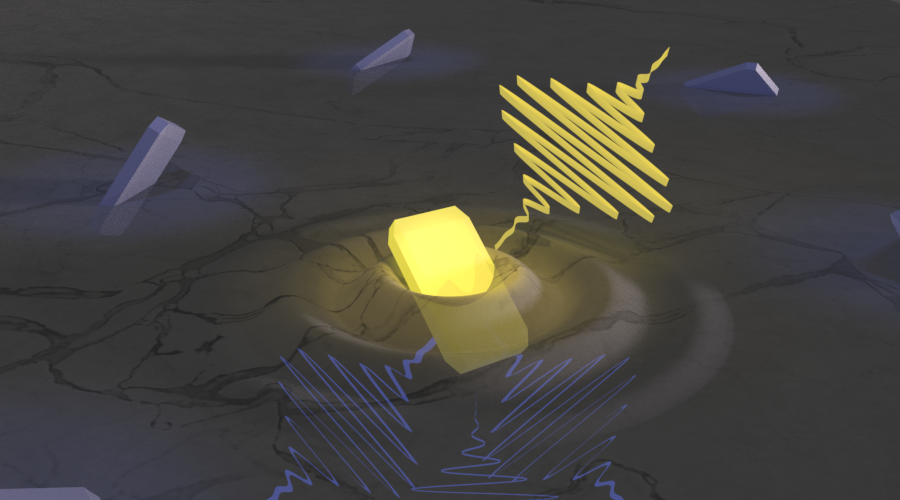}
    \end{tocentry}
\end{abstract}

Metal-halide perovskites are a class of materials that have attracted tremendous interest in recent years due to their superior performance in electronic transport and optical interactions \cite{Herz2016,Lozano2018,Zhao2019,Fu2019}. In particular, lead-halide perovskite nanocrystals, first synthesized in 2015 \cite{Protesescu2015}, have generated much excitement due to their unique light absorption and emission properties \cite{Akkerman2018,Shamsi2019}.

Although perovskite nanocrystals were initially limited to nanocube geometries, synthesis of perovskite nanoplatelets was achieved shortly thereafter \cite{Bekenstein2015,Tong2016}. Perovskite nanoplatelets share the efficient and tunable photo-physics of their nanocube counterparts while possessing unique characteristics of their own. Their planar geometry allows for directional light absorption/emission \cite{Jurow2019} and efficient energy transfer in stacked superlattices \cite{Rowland2015}, while precise control of the polyhedral layer thickness results in remarkable homogeneity in the dominant out-of-plane quantum confinement \cite{Weidman2016}. Because the absorption and emission energy of nanoplatelets is predominantly determined by the nanoplatelet thickness, variation in lateral confinement of electronic excitations is commonly believed to be negligible. 

However, many practical applications of colloidal nanoplatelets are still hampered by spectral broadening due to size and shape dispersion, known as inhomogeneous broadening. For example, overlap of absorption and emission bands of nanoplatelet ensembles degrades their efficiency as lasing media \cite{Li2015}. Inhomogeneous broadening also obscures the intrinsic homogeneous linewidth in absorption and luminescence measurements, a fundamental metric for many opto-electronic applications. In particular, elucidating the dominant homogeneous broadening mechanisms is crucial for optimizing energy transfer efficiencies of nanoplatelet superlattices \cite{Rowland2015,Weidman2016}. To properly determine the homogeneous and inhomogeneous linewidths and determine their underlying broadening mechanisms, a more advanced nonlinear spectroscopic technique is required to separate and characterize the two contributions to linewidth broadening. 

By correlating absorption and emission spectra, multi-dimensional coherent spectroscopy (MDCS) \cite{Cundiff2013} is able to separate homogeneous and inhomogeneous broadening mechanisms into orthogonal directions in a multi-dimensional spectrum. In particular, the ability of MDCS to circumvent inhomogeneous broadening has already proven invaluable in studying the homogeneous properties of colloidal nanoparticles \cite{Gellen2017,Seiler2018,Liu2019,Liu2019-2,Seiler2019,TripletExcitonPaper}. In this work, we apply MDCS at cryogenic temperatures to simultaneously determine the homogeneous and inhomogeneous linewidths of 3-layer and 4-layer CsPbI$_3$ perovskite nanoplatelet ensembles. We find that a change in thickness by a single layer drastically changes the degrees of both homogeneous and inhomogeneous broadening. Temperature- and excitation power-dependent measurements reveal the dominant linewidth broadening mechanisms to be acoustic phonon coupling and excitation induced dephasing (EID). By extrapolating their linear linewidth temperature dependences, we expect homogeneously broadened exciton resonances in 3-layer nanoplatelets at room-temperature, but not necessarily for thicker nanoplatelets.

\begin{figure}[b]
    \centering
    \includegraphics[width=0.5\textwidth]{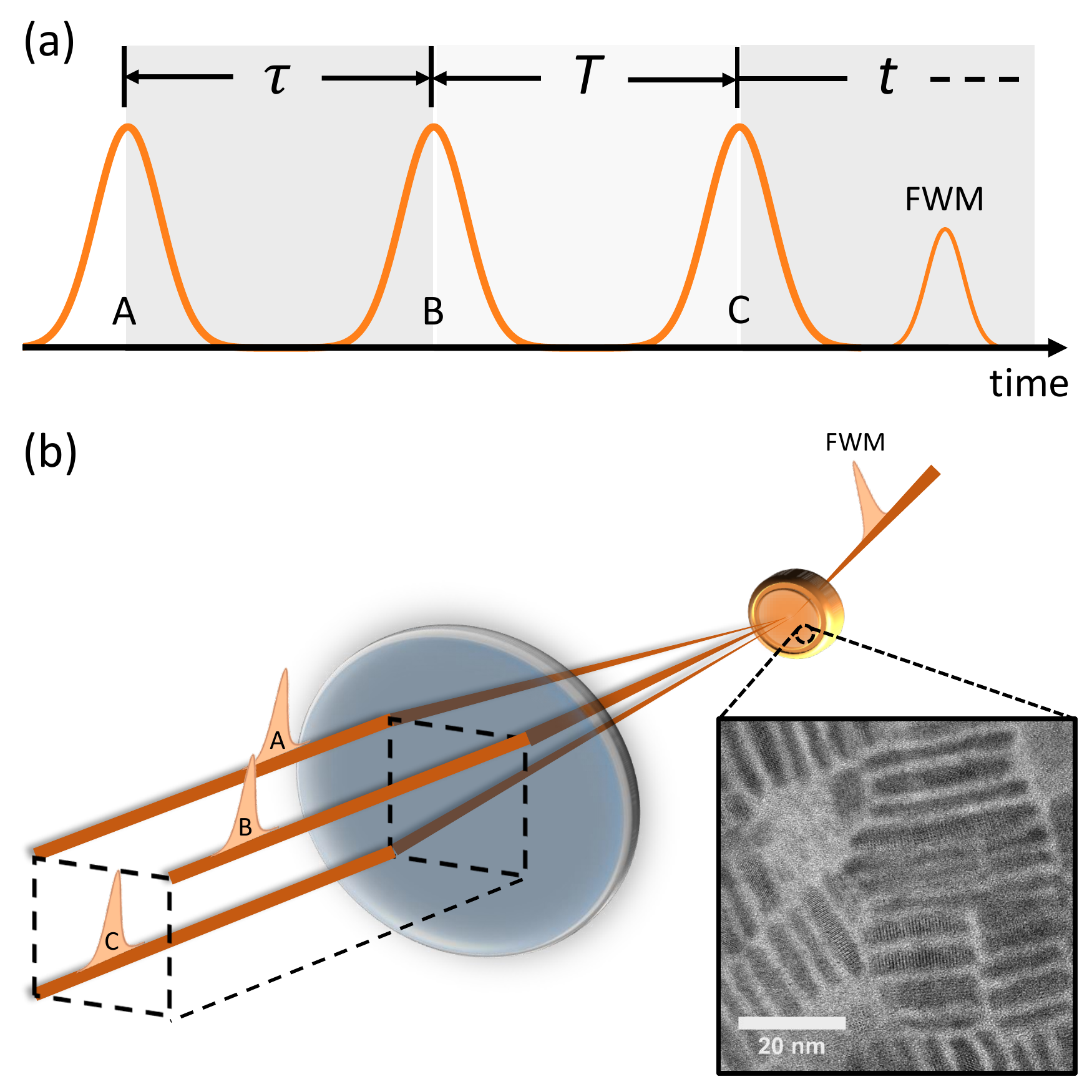}
    \caption{(a) Diagram defining the time-delays $\tau$, $T$, and $t$ that we Fourier transform along to generate an MDCS spectrum. (b) Experimental schematic, consisting of three phase-stable laser pulses arranged in the box geometry that generate a photon echo. Inset shows transmission electron microscopy image of studied 170$^\circ$C nanoplatelets.}
    \label{Fig1}
\end{figure}

\section{Results and Discussion}

To perform MDCS we use a Multi-Dimensional Optical Nonlinear Spectrometer \cite{Bristow2009}, in which three phase-stabilized laser pulses generate a nonlinear four-wave-mixing (FWM) signal as a function of three time delays $\tau$, $T$, and $t$ (shown in Fig.~\ref{Fig1}a). Fourier transforming the FWM signal as a function of delays $\tau$ and $t$ then generates a two-dimensional (2D) spectra (with conjugate axes $\hbar\omega_\tau$ and $\hbar\omega_t$) which correlates absorption and emission dynamics \cite{Liu2019,Liu2019-2}. 

In 2D spectra, inhomogeneous and homogeneous broadening manifest as broadening in the diagonal ($|\hbar\omega_\tau| = |\hbar\omega_t|$) and orthogonal cross-diagonal directions respectively \cite{Siemens2010}. We note that a first time-integrated FWM study has been performed on MAPbI$_3$ nanoplatelets to extract the homogeneous linewidth at a single temperature and excitation density \cite{Bohn2018}, but such measurements return a single ensemble-averaged value for the homogeneous linewidth and provide no information about its variation across the inhomogeneous distribution. In contrast, the cross-diagonal lineshape of a 2D spectrum at a given position along the diagonal $\hbar\omega_{\text{CD}} = |\hbar\omega_\tau| = |\hbar\omega_t|$ reflects the ensemble-averaged homogeneous response of nanoplatelets with a resonance energy $\hbar\omega_{\text{CD}}$. Fitting the diagonal and cross-diagonal lineshapes simultaneously \cite{Siemens2010} then provides the homogeneous and inhomogeneous linewidths ($\gamma = \frac{\hbar}{T_2}$ and $\sigma$ respectively). However, because the nanoplatelet absorption linewidths exceed the widths of our excitation spectra, the diagonal linewidths in our 2D spectra do not represent the total inhomogeneous distribution of nanoplatelet resonance energies. To prevent distortion of fitted homogeneous linewidths by such finite pulse bandwidth effects, we further divide out the experimental excitation spectrum for each 2D spectrum \cite{Smallwood2017,Do2017}.

\begin{figure}[b]
    \centering
    \includegraphics[width=0.5\textwidth]{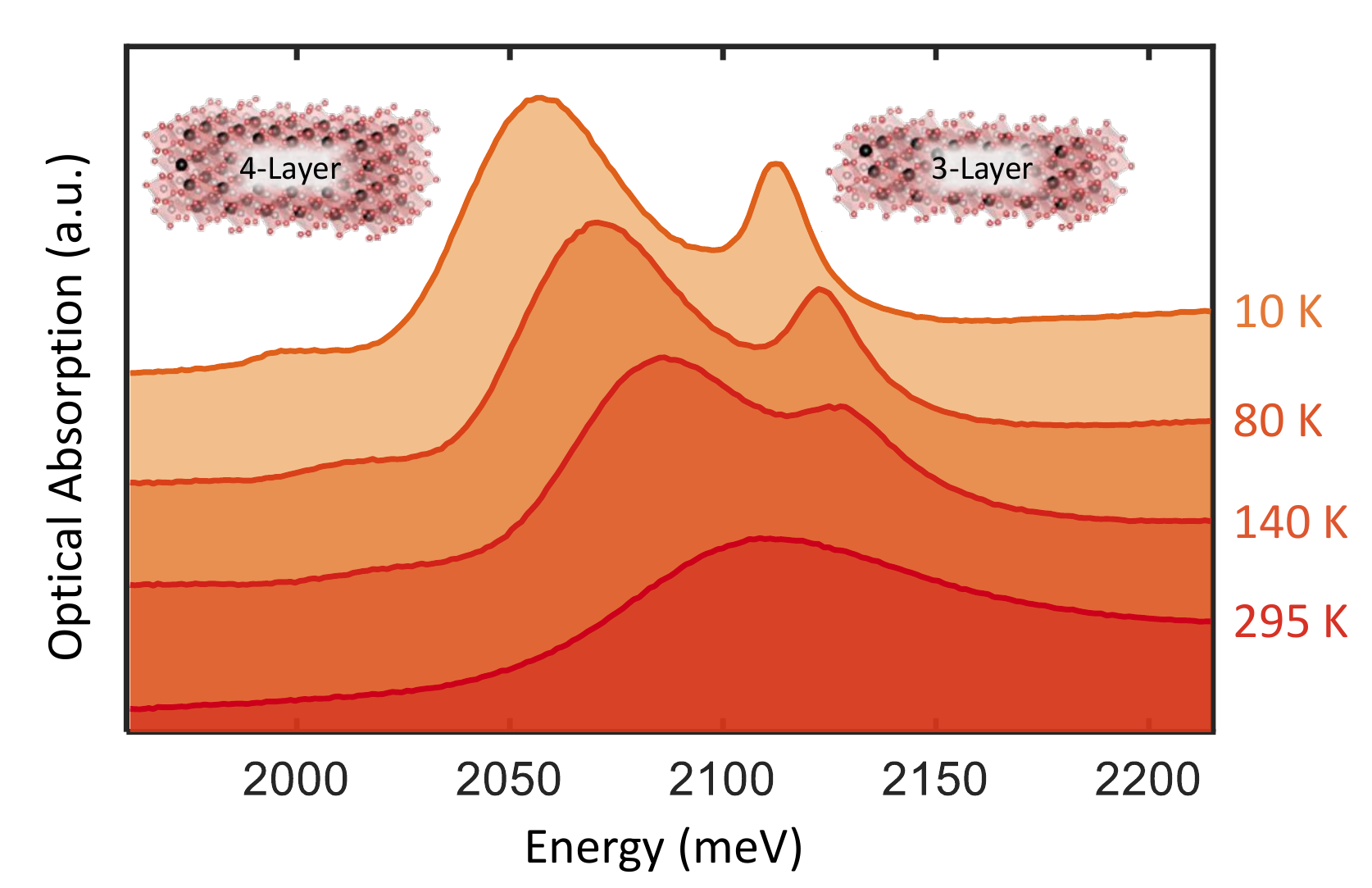}
    \caption{Temperature-dependent absorption spectra of nanoplatelets synthesized at 170$^\circ$C. A single broad absorption peak is visible at room-temperature, which splits into two distinct peaks attributed to different layer thicknesses at cryogenic temperatures.}
    \label{Fig2}
\end{figure}

Both 3-layer and 4-layer thick nanoplatelets were synthesized at a reaction temperature of 170$^\circ$C, as evident from the appearance of two distinct peaks in low-temperature absorption spectra shown in Fig.~\ref{Fig2}. Measurements were also performed on 4-layer nanoplatelets synthesized at a reaction temperature of 110$^\circ$C, at which no 3-layer nanoplatelets were synthesized. Transmission electron microscopy (TEM) measurements performed on each sample inform average lateral edge lengths of $10.7\pm 0.1$ nm and $12.4\pm 0.1$ nm for the 110$^\circ$C and 170$^\circ$C nanoplatelets, with size distributions of 4.6 nm and 4.0 nm respectively. No dependence of lateral size on nanoplatelet thickness was observed.

\begin{figure*}
    \centering
    \includegraphics[width=1\textwidth]{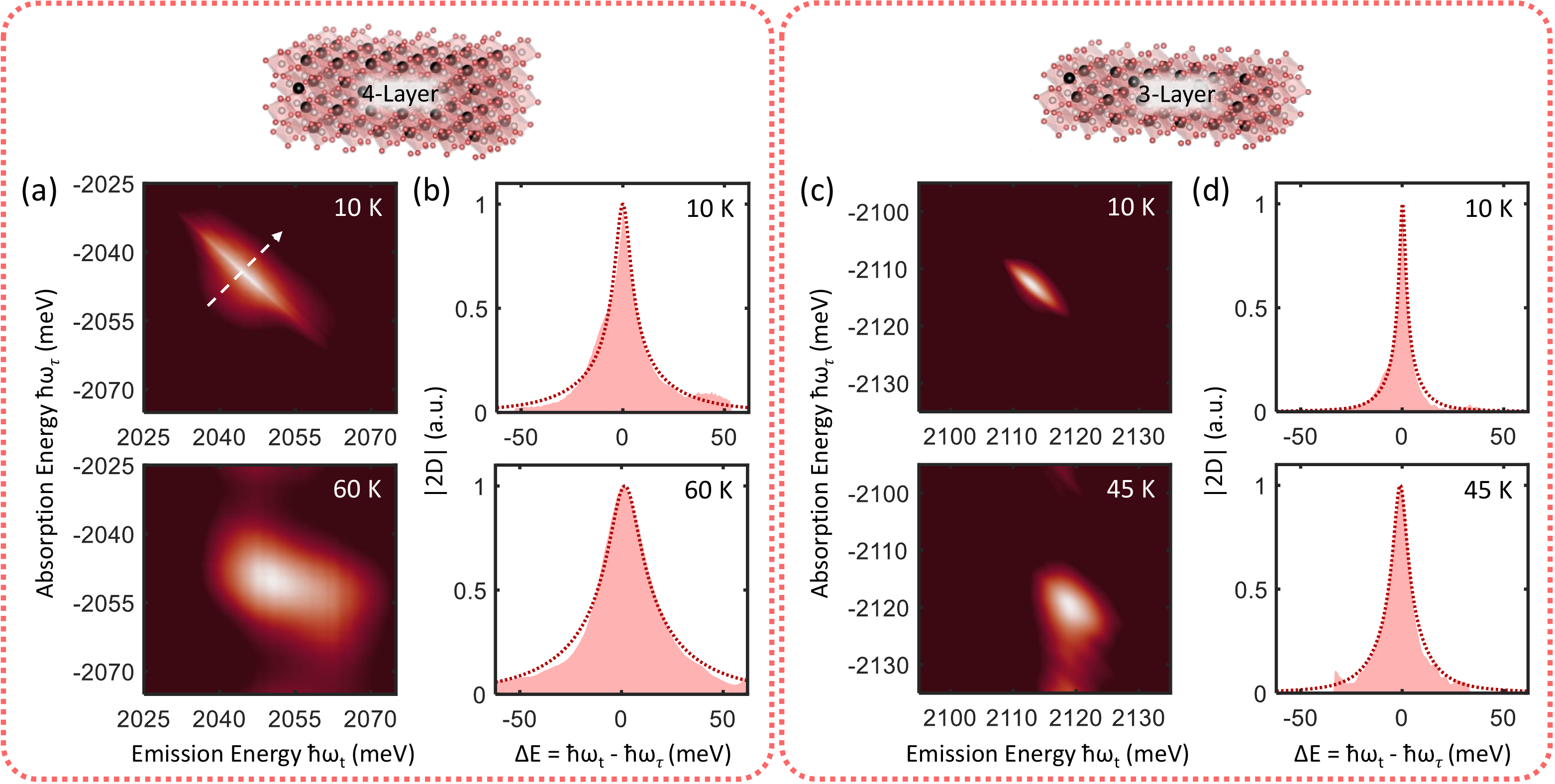}
    \caption{(a,c) Magnitude 2D spectra of (a) 4-layer and (c) 3-layer nanoplatelets synthesized at a reaction temperature of 170$^\circ$C and acquired with excitation densities $N_X = 1.753\times 10^{13}$ and $1.875\times 10^{13}$ cm$^{-2}$ respectively at the indicated temperatures. The waiting time was set to $T = 1$ ps to avoid coherent signals during pulse overlap. (b,d) Cross-diagonal slices of the (b) 4-layer and (d) 3-layer 2D spectra centered at 2045 meV and 2113 meV respectively. The cross-diagonal slice location for the 10 K slice in (b) is indicated by the white dashed arrow in (a). Experimental data and lineshape fits are plotted as the shaded area plots and dotted lines respectively.}
    \label{Fig3}
\end{figure*}

\subsection{Thermal Dephasing}

Magnitude 2D spectra of 4-layer and 3-layer nanoplatelets are shown in Figs.~\ref{Fig3}a and \ref{Fig3}c respectively. The vertical and horizontal axes represent the absorption and emission energies of the colloidal nanoplatelets as indicated, and the negative absorption energies reflect inverse phase evolution between delays $\tau$ and $t$. The diagonally-elongated lineshapes indicate that, despite dominant out-of-plane quantum confinement, exciton resonances in perovskite nanoplatelets still possess inhomogeneous broadening due to varying confinement of lateral exciton center-of-mass motion. The degree of inhomogeneous broadening depends strongly on layer thickness, which is evident from the difference in diagonal widths between Figs.~\ref{Fig3}a and \ref{Fig3}c.

With increasing temperature, the lineshapes in the cross-diagonal direction broaden, which is characteristic of thermal dephasing due to elastic exciton-phonon scattering \cite{Moody2015,Singh2013}. Specifically, first-order exciton-phonon scattering processes result in broadening that may be modelled by a linear temperature dependence for the homogeneous linewidth \cite{Rudin1990}:
\begin{align}\label{LinewidthTemperatureDependence}
    \gamma(T,N_{X}) = \gamma_0(N_{X}) + AT
\end{align}
where $\gamma_0(N_{X})$ is the zero-temperature linewidth at a given excitation density $N_{X}$ and the second term represents coupling to low-energy acoustic phonon modes with coupling strength $A$. In anticipation of the experimental linewidth temperature dependence, we neglect broadening due to discrete optical phonon modes that result in a nonlinear increase. To quantify thermal broadening in each system we plot cross-diagonal slices centered at $\hbar\omega_{\text{CD}} = 2045$ meV and 2113 meV in Figs.~\ref{Fig3}b and \ref{Fig3}d, which reflect homogeneous broadening of nanoplatelets with resonance energy $\hbar\omega_{\text{CD}}$. The lineshapes fit well to expressions derived for exponential dephasing in the Markovian limit \cite{Siemens2010}. As the nanoplatelet bandgaps blue-shift with increasing temperature, we adjust the slice locations by a commensurate energy.

\begin{figure*}
    \centering
    \includegraphics[width=1\textwidth]{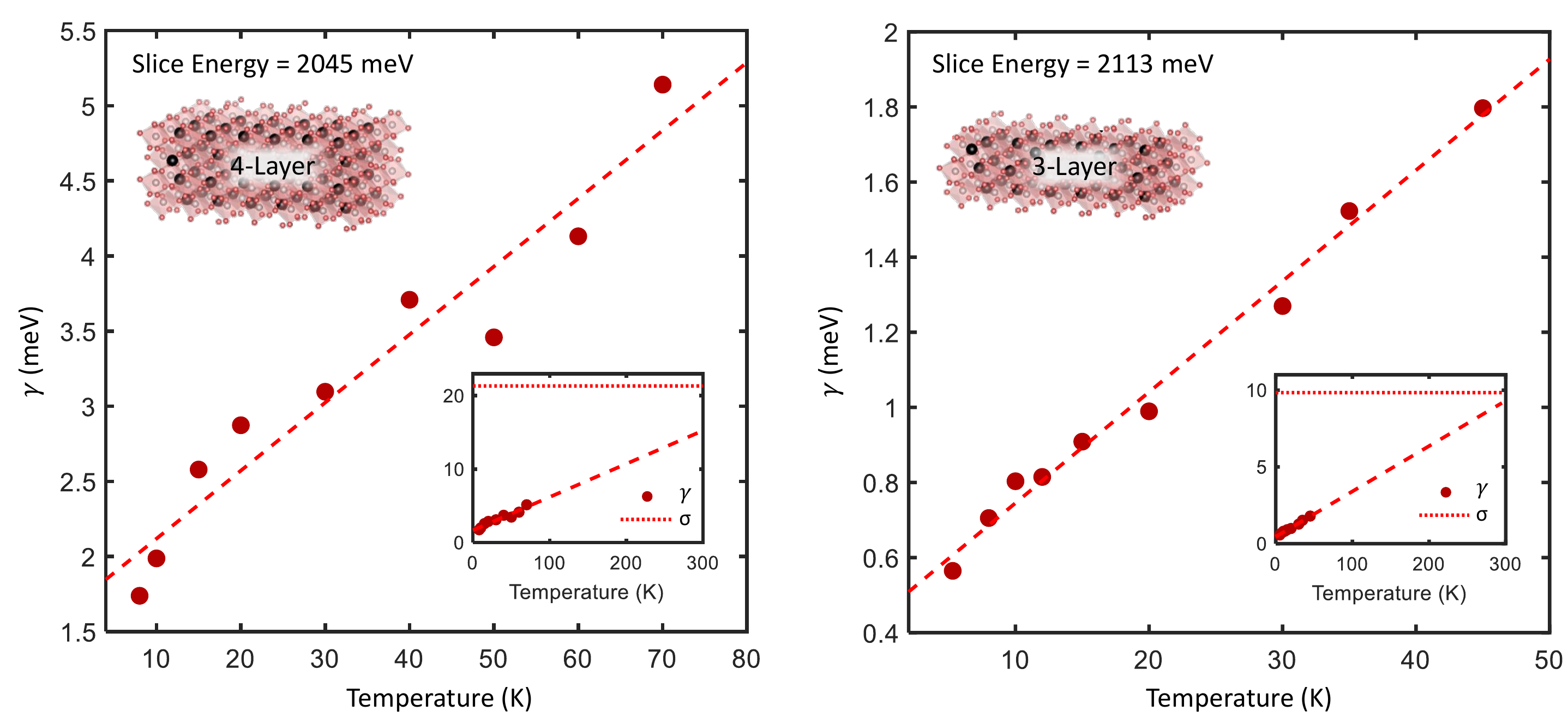}
    \caption{Dependence of the fitted homogeneous linewidths on temperature for 4-layer (left) and 3-layer (right) nanoplatelets synthesized at a reaction temperature of 170$^\circ$C. Experimental parameters are the same as in Fig.~\ref{Fig3}. The homogeneous linewidths for both samples follow linear temperature dependences. The linear fits shown are $\gamma = 1.67 + (0.045)T$ meV (4-layer) and $\gamma = 0.45 + (0.032)T$ meV (3-layer). The values of $\gamma$ and $\sigma$ extrapolated to room-temperature are plotted inset. The inhomogeneous linewidths, found by fitting the low-temperature absorption peaks to Voigt lineshapes as described in the text, are $\sigma = 21.32$ meV (4-layer) and $\sigma = 9.84$ meV (3-layer) and plotted as horizontal dotted lines in the insets. We note that the maximum 3-layer measurement temperature of 45 K is limited by a combination of the increase in resonance energy with temperature \cite{ThreeOscillatorPaper} and the tuning range of our excitation laser.}
    \label{Fig4}
\end{figure*}

The temperature dependent values of the homogeneous linewidth $\gamma$ (each fitted from a 2D spectrum taken at its respective temperature) are plotted in Fig.~\ref{Fig4}. By extrapolating to zero-temperature, we find drastically different intrinsic linewidths of $\gamma_0 = 0.45$ meV (3-layer) and $1.67$ meV (4-layer). To explain this surprising observation, we note that optical linewidths are fundamentally limited by the excited population relaxation rate. In colloidal nanocrystals, the dominant population relaxation channel is a spin-flip process between an optically-active bright-exciton manifold and an optically-inactive dark state which inhibits light-emission \cite{Accanto2012,Masia2012}. Decreasing nanoplatelet thickness (increasing electron-hole exchange interaction) results in a larger bright-dark energy splitting \cite{Shornikova2018}, and consequently a slower spin-flip relaxation rate. The linewidth also increases linearly with temperature, which is characteristic of acoustic phonon coupling as described by equation (\ref{LinewidthTemperatureDependence}). The thermal dephasing parameters extracted from the linear fits shown in Fig.~\ref{Fig4} are $A = 0.032$ meV/K (3-layer) and $A = 0.045$ meV/K (4-layer), which indicate relatively weaker acoustic phonon coupling in the former. Weaker vibrational coupling in 3-layer nanoplatelets is also reflected in their smaller thermal band-gap renormalization (shown in Fig.~\ref{Fig2}). Both values are comparable to that of similar two-dimensional systems such as monolayer WSe$_2$ (0.06 meV/K) \cite{Moody2015} and quantum wells ($\approx 0.01$ meV/K) \cite{Schultheis1986-2}.

Because optical phonon coupling is extremely weak in nanoplatelets, as evidenced by the linear bandgap temperature-dependence \cite{ThreeOscillatorPaper} and absence of vibrational sidebands in the experimental 2D spectra \cite{Liu2020}, it is reasonable to assume the homogeneous linewidth continues increasing linearly to higher temperatures. In the insets of Fig.~\ref{Fig4}, we plot the inhomogeneous linewidth (found by fitting the absorption peaks at 10 K to Voigt profiles \cite{Whiting1968} by using our respective measured values of $\gamma$) and extrapolate the homogeneous linewidths to room-temperature to predict the cross-over temperature at which homogeneous broadening exceeds inhomogeneous broadening. The crossover temperatures for 3-layer and 4-layer nanoplatelets are 292 K and 434 K, which indicate that thinner nanoplatelets are indeed homogeneously broadened at room temperature. In 4-layer nanoplatelets the extrapolated homogeneous linewidth at room temperature ($\gamma(\text{294 K}) = 14.98$ meV) is still significantly smaller than the inhomogeneous linewidth ($\sigma = 21.32$ meV), suggesting that further work is required to minimize size inhomogeneity in thick nanoplatelets for homogeneous exciton resonances at long-wavelengths.

\begin{figure}
    \centering
    \includegraphics[width=0.5\textwidth]{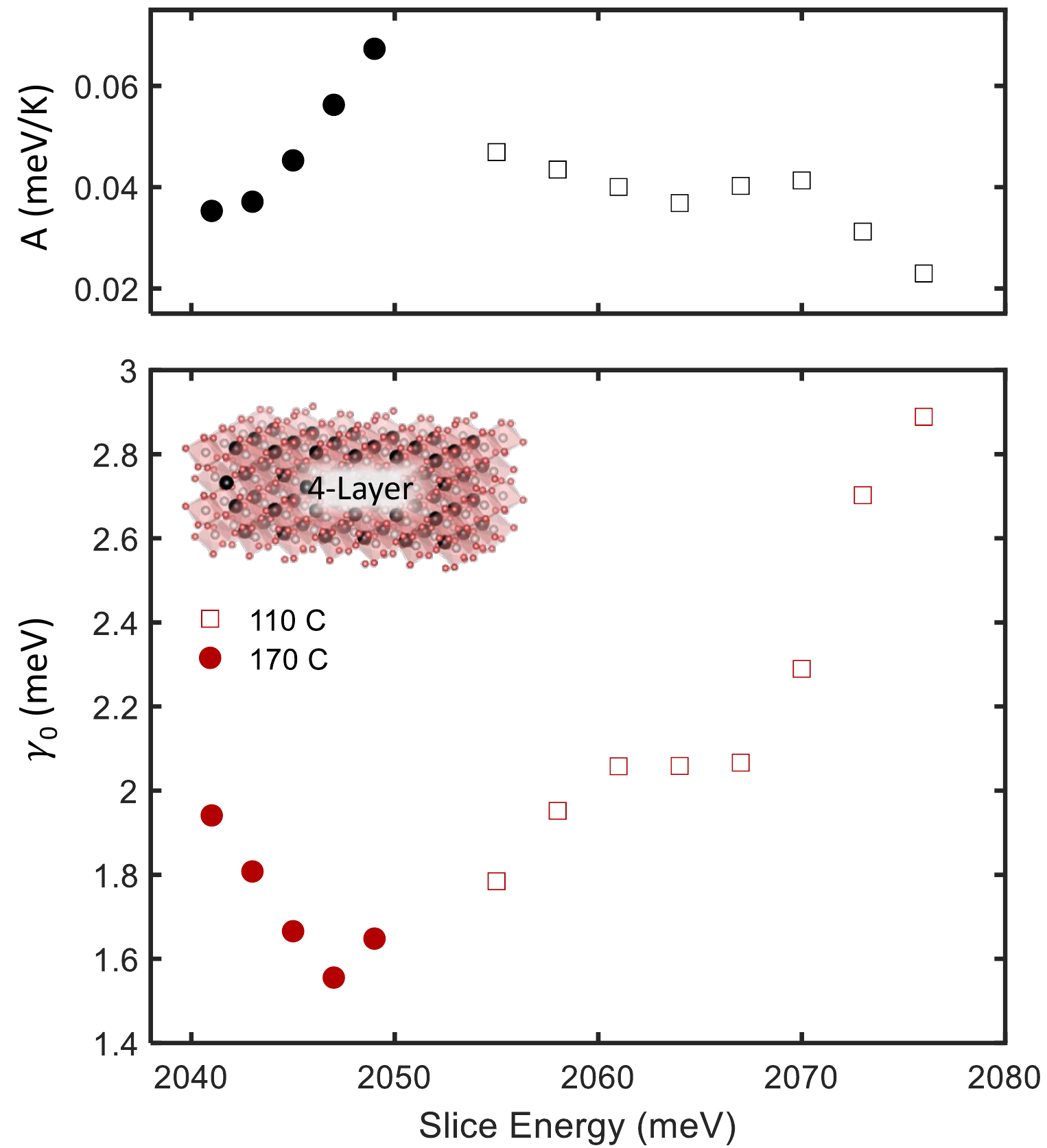}
    \caption{Dependence of the zero-temperature linewidth $\gamma_0$ (bottom plot) and thermal broadening parameter $A$ (top plot) on slice position for both 4-layer samples synthesized at 110$^\circ$C and 170$^\circ$C reaction temperatures. The values $\gamma_0$ are found by extrapolating the fitted linewidths to zero-temperature, as shown in Fig.~\ref{Fig4}.}
    \label{Fig5}
\end{figure}

The large inhomogeneous broadening in 4-layer nanoplatelets also invites further resolving their temperature dependence analysis in terms of resonance energy, which corresponds to the cross-diagonal slice position in a 2D spectrum. In colloidal nanoplatelets, the resonance energy distribution directly corresponds to variations in lateral size and quantum confinement. The extrapolated zero-temperature homogeneous linewidth $\gamma_0$ is plotted in Fig.~\ref{Fig5} as a function of slice position, which exhibits an increase with increasing slice position (resonance energy). The slice position (resonance energy) dependence of the thermal dephasing parameter $A$ is then plotted inset for both samples in Fig.~\ref{Fig5} which appear to vary inversely with $\gamma_0$.

\subsection{Excitation-Induced Dephasing}

Another dominant extrinsic dephasing mechanism in semiconductors is excitation-induced dephasing (EID) \cite{Wang1994,Shacklette2002,Moody2015}, which arises from electronic many-body interactions. Our narrow excitation bandwidths, well-matched to the probed exciton resonances, restricts the source of EID to exciton-exciton scattering \cite{Schultheis1986,Moody2015}. In two-dimensional systems, this may be described by a linear dependence on excitation density $N_X$ for the homogeneous linewidth:
\begin{align}\label{LinewidthPowerDependence}
    \gamma(T,N_X) = \gamma(T) + BN_X
\end{align}
where $\gamma(T)$ is the zero-density linewidth at temperature $T$ and $B$ is the exciton-exciton interaction strength. The excitation densities are calculated from the experimental laser parameters and sample optical density.

\begin{figure}[b]
    \centering
    \includegraphics[width=0.5\textwidth]{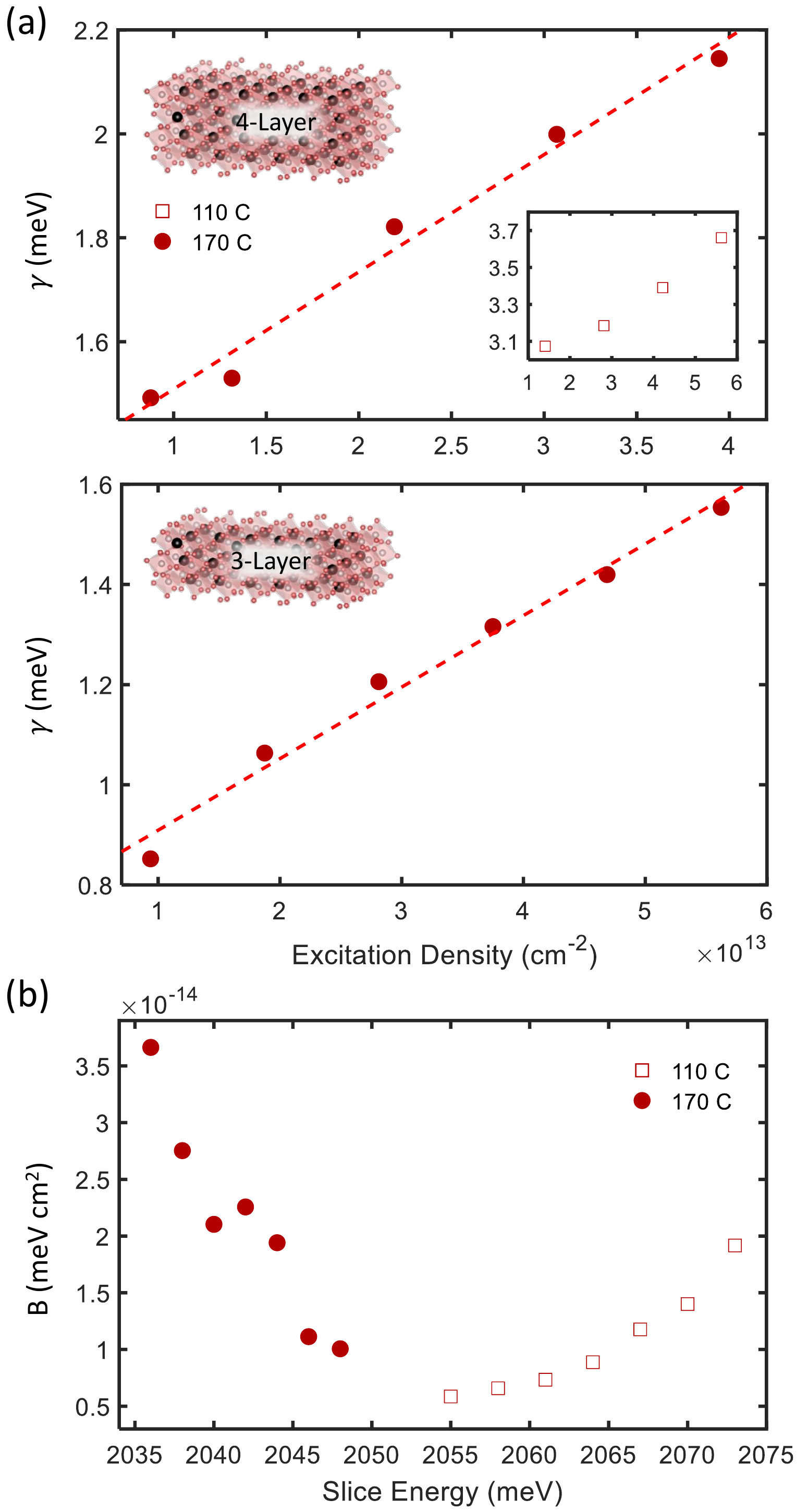}
    \caption{(a) Fitted values of the homogeneous linewidth as a function of excitation density for 4-layer (top) and 3-layer (bottom) nanoplatelets at 10 K. The 4-layer linewidths are fitted from slices taken at 2070 meV (110$^\circ$C, inset) and 2042 meV (170$^\circ$C) while the 3-layer linewidths are fitted from slices at 2113 meV. (b) Dependence of the EID parameter $B$ on slice position in 4-layer nanoplatelets.}
    \label{Fig6}
\end{figure}

The homogeneous linewidth at 10 K is plotted in Fig.~\ref{Fig6}a as a function of excitation density for both 3-layer and 4-layer nanoplatelets. We find that the linewidth increases linearly with excitation density for both thicknesses and reaction temperatures, a clear signature of EID, as described by equation (\ref{LinewidthPowerDependence}). The EID parameter $B$ for 4-layer nanoplatelets synthesized at both temperatures is then plotted as a function of slice location in Fig.~\ref{Fig6}b, which, most notably, exhibits a rapid increase in EID below 2055 meV. The sharp increase in EID with decreasing resonance energy for 170$^\circ$C nanoplatelets may be understood by noting the linear dependence of absorption cross-section on lateral area \cite{Yeltik2015} and by considering the effect of nanoplatelet size on multiple exciton dynamics. In large nanoplatelets (smaller resonance energy) multiple excitons may form, which may then undergo scattering events that result in EID. Generation of multiple excitons becomes favorable once the nanoplatelet size exceeds the exciton Bohr diameter in CsPbI$_3$ of 12 nm \cite{Protesescu2015}, which agrees with the mean edge length of 12.4 nm measured for the 170$^\circ$C nanoplatelets. The observed increase in EID beyond 2055 meV (with decreasing nanoplatelet size) is unexpected, which we speculate as due to polaronic effects that result in increasing sensitivity to EID \cite{Roy2011}. Because the nanoplatelet size is still comparable to the exciton Bohr radius, the EID parameters measured here are much smaller than the EID parameter in, for example, monolayer WSe$_2$ ($2.7\times 10^{-12}$ meV~cm$^2$ \cite{Moody2015}). We note that the EID parameter measured for 3-layer nanoplatelets ($B = 1.43\times 10^{-14}$ meV~cm$^2$) lies between the range of values measured for 4-layer nanoplatelets synthesized at the same temperature, which indicates minimal effect of nanoplatelet thickness on EID in this size regime.

\section{Conclusions}

In conclusion, we have measured the intrinsic homogeneous linewidth and its broadening mechanisms in CsPbI$_3$ perovskite nanoplatelets as a function of platelet geometry. By informing the relevant design parameters for optical resonances, our results are directly relevant for implementing colloidal nanoplatelets in practical devices. For example, we find that large-area nanoplatelets offer advantages such as narrow intrinsic homogeneous linewidths and lower thermal broadening, and are therefore likely to be advantageous in coherent opto-electronic devices. However, small-area nanoplatelets (that suppress multiple exciton formation) may be required for narrow and stable exciton resonances in high optical intensity applications. We realize the narrowest, sub-meV intrinsic linewidths by reducing the nanoplatelet thickness to a mere three polyhedral layers, which represents an important step towards engineering ideal atom-like emission from colloidal materials.

\section{Methods}

\subsection{CsPbI$_3$ Nanoplatelet Synthesis Method}

CsPbI$_3$ nanoplatelets were synthesized according to a procedure detailed elsewhere \cite{Bonato2020}. Briefly, 15 mg of Cs$_2$CO$_3$ and 158 mg of Pb(acetate)$_2$3H$_2$O were loaded into 50 mL, 3-neck flask along with 1 mL of oleic acid (OA), 0.5 mL of oleylamine (OLA) and 4.5 mL of 1-octadecene (1-ODE). The mixture was kept at 100$^\circ$C for 1 hour under reduced pressure. After complete degassing of the mixture, the system was kept under nitrogen flowing and heated-up until the specified reaction temperature (either 110$^\circ$C or 170$^\circ$C). Then, a solution containing 130 mg of SnI$_4$ in 1 mL of 1-ODE, 0.5 mL of OA and 0.5 mL of OLA (all solvents were pre-degassed before using) was swiftly injected. The mixture was rapidly placed into a water bath until room temperature. The nanoplatelets were separated from the crude solution by centrifugation at 13,500 rpm for 10 minutes and re-suspended with anhydrous n-hexane. 

\subsection{Cryogenic Temperature Measurements}

The CsPbI$_3$ nanoplatelets are suspended in heptamethylnonane, a branched alkane that forms a transparent glass at temperatures below 100 K. The colloidal suspension is held between two sapphire windows in a custom sample holder approximately 0.5 mm thick, which is then mounted to a liquid helium cold-finger cryostat and cooled to cryogenic temperatures.

\subsection{MDCS Measurements}

The excitation laser pulses used to perform multi-dimensional coherent spectroscopy (MDCS) are generated at a 250 kHz repetition rate by an optical parameteric amplifier (Coherent OPA 9400) pumped by a regenerative Ti:Sapphire amplifier (Coherent RegA 9000). The FWM signal is isolated by wave-vector phase-matching in the box geometry (shown in Fig.~\ref{Fig1}b), measured via spectral interferometry \cite{Lepetit1995} with a fourth phase-stabilized local oscillator pulse. All pulses are co-linearly polarized and measured to be approximately 70 fs in duration via intensity autocorrelation.

\subsection{Exciton Density Calculation}

To calculate the exciton excitation density, we use:
\begin{equation}
    N_X = \frac{P_{\text{avg}}(1 - 10^{-\text{OD}})}{\pi r^2f_{\text{rep}}E_{\text{ph}}}
\end{equation}
where $P_{\text{avg}}$ is the average power per excitation beam, $A = 1 - 10^{-\text{OD}}$ is the linear absorbance of each respective nanoplatelet ensemble, $r = 55\mu$m is the focused beam radius at the sample, $f_{\text{rep}} = 250$ kHz is the excitation repetition rate, and $E_{\text{ph}}$ is the excitation photon energy.

\begin{acknowledgement}

We acknowledge M. Siemens for fruitful discussions. This work was supported by the Department of Energy grant number DE-SC0015782. G.N. and D.B.A. acknowledge support by fellowships from the Brazilian National Council for Scientific and Technological Development (CNPq). L.A.P. acknowledges support from FAPESP (Project numbers 2013/16911-2 and 2016/50011-7). Research was supported by LNNano/CNPEM/MCTIC, where the TEM measurements were performed. 

\end{acknowledgement}

\bibliography{bibliography}

\end{document}